\documentclass[preprint,showpacs,showkeys,preprintnumbers,amsmath,amssymb,aps,pra]{revtex4-1}

\usepackage{bm,amssymb,amsmath,mathrsfs}
\usepackage[usenames]{color}
\usepackage{enumerate}
\usepackage{hyperref}
\usepackage{bookmark}

\usepackage{dcolumn}
\usepackage[pdftex]{graphicx}

\def\half{{\textstyle\frac{1}{2}}}

\newcommand{\bc}{\begin{center}}
\newcommand{\ec}{\end{center}}
\newcommand{\bit}{\begin{itemize}}
\newcommand{\eit}{\end{itemize}}
\newcommand{\bq}{\begin{equation}}
\newcommand{\eq}{\end{equation}}

\begin{document}

\preprint{CC 15\_10}

\title{Spin Wheel}

\author{S.~R.~Mane}
\email{srmane001@gmail.com}

\affiliation{Convergent Computing Inc., P.~O.~Box 561, Shoreham, NY 11786, USA}

\date{\today}

\begin{abstract}
\bc
A distillation of Koop's idea of the Spin Wheel.
\ec
\end{abstract}



\maketitle


\baselineskip 4ex

\section{\label{sec:intro} Introduction}
Recently, I had the pleasure to read an excellent paper by Ivan Koop, 
presenting an idea which he has termed the `Spin Wheel,'
to detect a possible electric dipole moment (EDM) for unbound charged particles.
Koop has presented the idea of the Spin Wheel at various conferences 
\cite{Koop_IPAC_2013}
and talks
\cite{Koop_Juelich_2012},
but of necessity these presentations contain much additional material.
This note is a distillation of the core idea of the Spin Wheel, pure and simple,
a summary devoid of clutter, as far as I can make it so.
I thank Ivan Koop for his kind permission to publish this note.

\section{\label{sec:ham} Dipole Moments and Hamiltonian}
In a rigorous treatment, the electric and magnetic dipole moments are the coefficients of operators using the Dirac equation.
However, the Dirac equation applies only for particles of spin $\frac12$,
and we wish to treat particles such as deuterons, which have spin 1.
The treatment below begins with a particle at rest, followed by a semiclassical treatment of relativistic motion.
Let the particle spin be $\bm{s}$ and define $\bm{S}=\bm{s}/s$, where
$s=\frac12\hbar,\hbar,\dots$ for particles of spin $\half,1,\dots$, etc.
The spin Hamiltonian is
\bq
H_{\rm rest} = -(\mu \bm{B} + d\bm{E})\cdot\bm{S} \,.
\eq
Here $\bm{E}$ and $\bm{B}$ are the external electric and magnetic fields, respectively,
and $\mu$ is the magnetic dipole moment (MDM) and $d$ is the electric dipole moment (EDM).
To generalize to relativistic motion,
we retain $\bm{S}$ as the (dimensionless) spin operator in the rest frame,
but $\bm{E}$ and $\bm{B}$ are the external electric and magnetic fields in the laboratory frame.
Let the particle have velocity $\bm{v}=\bm{\beta}c$ and Lorentz factor $\gamma = 1/\sqrt{1-\beta^2}$.
The fact that we employ $\bm{\beta}$ and $\gamma$ already indicates a semiclassical treatment of the orbital motion.
Thomas precession also appears in the Hamiltonian, which is
\bq
\begin{split}
\label{eq:hrel}
H_{\rm rel} &= -\frac{1}{\gamma}(\mu \bm{B}^\prime + d\bm{E}^\prime)\cdot\bm{S} + \bm{\omega}_T\cdot\bm{S} 
\equiv (\bm{\Omega}_{MDM} + \bm{\Omega}_{EDM})\cdot\bm{S} \,.
\end{split}
\eq
The factor of $1/\gamma$ is due to time dilation.
Here $\bm{E}^\prime$ and $\bm{B}^\prime$ are the electric and magnetic fields in the rest frame
and $\bm{\omega}_T$ is the Thomas precession vector and with an obvious notation
$\bm{\Omega}_{MDM}$ and $\bm{\Omega}_{EDM}$ are the spin precession vectors for the MDM and EDM respectively,
where $\bm{\Omega}_{MDM}$ includes the contribution of the Thomas precession. Then
\begin{subequations}
\begin{align}
\bm{\Omega}_{MDM} &= -\frac{e\hbar}{mc}\biggl[\biggl(a+\frac{1}{\gamma}\biggr)\bm{B}
-\frac{a\gamma}{\gamma+1}\,\bm{\beta}\cdot\bm{B}\bm{\beta}
-\biggl(a+\frac{1}{\gamma+1}\biggr)\bm{\beta}\times\bm{E} \biggr] \,,
\\
\bm{\Omega}_{EDM} &= -\frac{e\hbar\eta}{mc}
\Bigl(\bm{E} + \bm{\beta}\times\bm{B} - \frac{\gamma}{\gamma+1}\, \bm{\beta}\cdot\bm{E}\,\bm{\beta} \Bigr) \,.
\end{align}
\end{subequations}
The particle mass is $m$, the charge is $e$, $a=(g-2)/2$ is the magnetic moment anomaly and we write
$d = e\hbar\eta/(mc)$ for the electric dipole moment.
In this note, I shall treat only motion on the reference orbit,
hence $\bm{\beta}\cdot\bm{E} = \bm{\beta}\cdot\bm{B}=0$.

\section{\label{sec:neutron} Neutrons}
Most EDM experiments to date have been performed with electrically neutral systems,
such as neutrons or else electrons bound in atoms or molecules (which are overall neutral)
or nuclei such as ${}^{199}$Hg.
There are important lessons here, which directly motivate the idea of the Spin Wheel.
I shall briefly discuss EDM experiments using neutrons.

A batch of vertically polarized neutrons is introduced into a sample cell.
The electric and magnetic fields $\bm{E} = E_z\bm{e}_z$ and $\bm{B}=B_z\bm{e}_z$ are both vertical (and uniform).
A $\frac{\pi}{2}$ rf pulse is applied to rotate the neutron spins into the horizontal plane.
The neutron spins then preccess freely in the horizontal plane.
This is done for multiple batches, but the direction of the electric field is reversed.
The spin precession frequencies in the various runs are measured
\bq
\label{eq:nedm}
\omega_\pm = \frac{\mu B_z \pm dE_z}{\hbar} \,.
\eq
The frequency difference $\omega_+ - \omega_-$ is proportional to the electric dipole moment 
\bq
d = \frac{\hbar(\omega_+ - \omega_-)}{2E_z} \,.
\eq
Note the following important points:
\bit
\item
Significantly, the experiment consists of measuring {\em frequencies} only.
It is well known that frequencies can be experimentally measured with great precision.
\item
The applied electric and magnetic fields are {\em parallel}.
Hence the spin precession frequency is {\em linear} in the electric dipole moment.
\item
To enhance the EDM signal, the applied electric field $E_z$ is strong.
However, because the neutrons have no electric charge, they experience no Lorentz force 
and hence do {\em not} accelerate out of the sample cell when an external electric field is applied.
\eit
The above points are all fundamental to the notion of the Spin Wheel.

\section{\label{sec:free} Unbound charged particles}
\subsection{General remarks}
For atomic physics EDM experiments for electrons bound in atoms or molecules,
the applied electric and magnetic fields are also parallel
and the differences between the electron energy levels (which can be measured as frequencies) 
are also linear in the electric dipole moment.
Because the system is overall electrically neutral, 
the electrons remain in their molecular orbitals 
when a strong external electric field is applied.
The situation is markedly different for unbound charged particles.
The application of a strong external electric field will cause unbound charged particles to accelerate out of the sample cell.
The solution which has been proposed, to work around this problem, 
is to circulate a beam of spin polarized charged particles in a storage ring.
The external electric and magnetic fields then not only cause the spins to precess,
but also constrain the particles to a circulate in a storage ring.

Let us employ cylindrical polar coordinates $(x,y,z)$, 
where $\bm{e}_x$ is radial (outwards),
$\bm{e}_y$ is tangential (along the reference orbit),
and $\bm{e}_z$ is vertical (upwards).
I shall also write $\bm{e}_r=\bm{e}_x$ for the radial unit vector.
The positive sense of circulation is counterclockwise.
In the simplest model of a ring, 
the external electric field is radial $\bm{E}=E_r\bm{e}_r$ and
the external magnetic field is vertical $\bm{B}=B_z\bm{e}_z$.
It is readily seen that the Lorentz force $e(\bm{E}+\bm{\beta}\times\bm{B})$ is radial on the reference orbit.
It is also readily seen that 
$\bm{\Omega}_{MDM} \parallel \bm{e}_z$ is vertical and
$\bm{\Omega}_{EDM} \parallel \bm{e}_r$ is radial.
This leads to an immediate difficulty if we attempt to measure the spin precession frequency:
\bq
\label{eq:omnaive}
\omega \propto \sqrt{\bm{\Omega}_{MDM}^2 + \bm{\Omega}_{EDM}^2}
\simeq \bm{\Omega}_{MDM} + \frac{\bm{\Omega}_{EDM}^2}{2\bm{\Omega}_{MDM}} \,.
\eq
Because $\bm{\Omega}_{MDM}$ and $\bm{\Omega}_{EDM}$ are {\em orthogonal}, 
the frequency shift is of {\em second order} in the electric dipole moment.
This greatly reduces the sensitivity of the experiment.

As a matter of fact, the above model is exactly how all experiments to detect the 
electric dipole moment of the muon have been carried out to date.
In all the experiments to measure the muon $g-2$ (anomalous magnetic dipole moment),
the muons were circulated in a (highly uniform) vertical magnetic field,
hence $\bm{\Omega}_{MDM} \parallel \bm{e}_z$ was vertical.
There was no strong radial electric field 
(although electrostatic quadrupole focusing was employed in some muon $g-2$ rings)
but nevertheless $\bm{\Omega}_{EDM} \parallel \bm{v}\times\bm{B} \parallel \bm{e}_r$ was radial.
Hence if the muons possess a nonzero electric dipole moment, the muon polarization should precess out of the horizontal plane.
No such precession was detected, which sets the currently known limits on the electric dipole moment of the muon.
However, the limits for the muon electric dipole moment are much less precise than those for the neutron and electron.

\subsection{Helicity and the Frozen Spin Method}
We therefore seek ideas to work around the limitation of eq.~\eqref{eq:omnaive}.
Since the electric field must be radial (to confine the particles in a ring), $\bm{\Omega}_{EDM}$ must be radial.
Hence we must manipulate $\bm{\Omega}_{MDM}$.
This is done in two steps.
The first step is to consider the equation of motion for the helicity $\bm{S}\cdot\hat{\bm{\beta}}$.
Neglecting the EDM term, which is small, the equation is
\bq
\label{eq:eqhel}
\frac{d\ }{dt} (\bm{S}\cdot\hat{\bm{\beta}}) = 
-\frac{e}{mc}\biggl[\,a\bm{B} -\biggl(a - \frac{1}{\beta^2\gamma^2}\biggr)\bm{\beta}\times\bm{E}\biggr]\cdot
(\bm{S}\times\hat{\bm{\beta}}) \,.
\eq
Suppose we choose the fields so that the term in brackets {\em vanishes}.
In this case the helicity will be an invariant of the motion.
Note that eq.~\eqref{eq:eqhel} is general, and independent of any coordinate system.
For our storage ring model, we set
\bq
aB_z +\biggl(a - \frac{1}{\beta^2\gamma^2}\biggr)\beta E_r = 0 \,.
\eq
This idea is called the `frozen spin method.'
If a spin is initially longitudinal, i.e.~the helicity is $+1$, it will remain so.
In general, for a set of spins in the horizontal plane in the ring,
when the spins are viewed in the `accelerator frame,'
the spins will appear to be `frozen' in place and not precess.
The effect of $\bm{\Omega}_{MDM}$ on the spins has effectively been nulled.

Of course there is also $\bm{\Omega}_{EDM}$, which is radial.
If the polarization vector is initially $100\%$ along the beam axis,
then $\bm{\Omega}_{EDM}$ will cause the polarization vector to rotate around the radial axis and develop a nonzero vertical component.
One design of an EDM storage ring is to attempt to detect such a nonzero vertical component of the beam polarization.
The estimated rotation angle of the polarization vector,
at the limit of sensitivity of the experiment, is about a microradian.
This is a challenging task for the polarimetry to detect.

\subsection{Spin Wheel}
Here is where Koop introduced his idea of the `Spin Wheel' \cite{Koop_IPAC_2013,Koop_Juelich_2012}.
First we impose the above frozen spin condition, to fix the value of the vertical magnetic field $B_z$.
As we saw above, in the accelerator frame, this effectively makes $\bm{\Omega}_{MDM}=0$.
In addition, we now introduce a {\em radial} magnetic field $B_r$.
Note that this magnetic field is {\em parallel} to the electric field $E_r$.
Then, in the accelerator frame, we now have an effective radial MDM spin precession vector 
$\bm{\Omega}_{MDM}^\prime \parallel \bm{e}_r$.
Koop's Spin Wheel idea therefore yields a storage ring configuration where 
$\bm{\Omega}_{MDM}^\prime$ and $\bm{\Omega}_{EDM}$ are parallel.
The EDM signal is therefore linear in the spin precession frequency.
When viewed in the `accelerator frame,' the spins precess around the radial axis with a frequency
\bq
\tilde{\omega}_\pm = \frac{\pm \Omega_{MDM}^\prime + \Omega_{EDM}}{\hbar} \,.
\eq
I affix the tilde to distinguish this from eq.~\eqref{eq:nedm} for the neutron EDM spin precession frequency,
and also to indicate that this frequency is for motion in the accelerator frame,
i.e.~spin precession around the radial axis.
Also, in a storage ring, we cannot reverse the direction of the electric field $E_r$,
hence we must reverse the direction of the radial magnetic field $B_r$, i.e.~$\Omega_{MDM}^\prime$.
Note that, as with the neutron EDM expeiments, the experiment consists of pure frequency measurements.
The electric dipole moment is obtained via the frequency {\em sum}
\bq
d \propto \frac{\tilde{\omega}_+ + \tilde{\omega}_-}{2} \,.
\eq
Note also that a radial magnetic field $B_r$ will deflect the particles vertically,
via a $e(\bm{\beta}\times\bm{B})$ Lorentz force,
and will also generate a vertical term in $\bm{\Omega}_{EDM}$.
This $e(\bm{\beta}\times\bm{B})$ force will be compensated by the vertical focusing gradients in the ring.
Hence the beam centroid will be displaced vertically from the reference orbit,
and the contribution to $\bm{\Omega}_{EDM}$ will be cancelled.
A possible idea to avoid such a vertical deflection is to employ a Wien filter.
A Wien filter is a pair of crossed electric and magnetic fields such that $\bm{E} +\bm{\beta}\times\bm{B}=0$,
so the Lorentz force on the orbit vanishes.
In the present context, this requires a {\em vertical} electric field $E_z$ such that $E_z -\beta B_r=0$.
Note that because $\bm{E} +\bm{\beta}\times\bm{B}=0$
(and also $\bm{\beta}\cdot\bm{E}=0$),
there is no contribution to $\bm{\Omega}_{EDM}$ from the Wien filter,
so $\bm{\Omega}_{EDM}$ is still radial.
The overall set of guide fields in the ring is 
$(B_z,E_r)$, which are linked by the frozen spin condition, and
$(B_r,E_z)$, which are linked by the Wien filter condition.
(In addition, there are focusing fields to focus the particle orbits.)



\end{document}